# Switching magnetic vortex core by a single nanosecond current pulse

Keisuke Yamada[1], Shinya Kasai[1], Yoshinobu Nakatani[2], Kensuke Kobayashi[1] & Teruo Ono[1]

[1]*Institute for Chemical Research, Kyoto Univeristy, Uji 611-0011, Japan*

[2]*University of Electro-communications, Chofu 182-8585, Japan*

**Abstract**

In a ferromagnetic nanodisk, the magnetization tends to swirl around in the plane of the disk and can point either up or down at the center of this 'magnetic vortex'. This binary state can be useful for information storage. It is demonstrated that a single nanosecond current pulse can switch the core polarity. This method also provides the precise control of the core direction, which constitutes fundamental technology for realizing a vortex core memory.

Since the experimental confirmation of the existence of a nanometer-scale core with out-of-plane magnetization in a magnetic vortex,[1, 2] the dynamics of a vortex has been investigated intensively,[3-11] and the magnetic vortex core has been expecting to be a candidate for future nonvolatile data storage devices because of its small size and its good thermal stability. However, flipping the vortex core direction from up to down requires a fairly large magnetic field.[12] Recently, this technological hurdle has been reduced by utilizing the resonance effect of the vortex core motion induced either by the application of an ac magnetic field[13, 14] or by injecting an ac electric current into a magnetic disk.[9, 15] We have shown that the core switches when its velocity reaches a certain value regardless of the value of the excitation current density, indicating that the core switching is triggered by a strong dynamic field which is produced locally by a rotational core motion at a high speed of several hundred m/s in either case of the magnetic field or electric current excitation.[15] Although the final core direction can not be controlled in this method, this understanding leads to the idea that the core can be switched if the core is accelerated to high enough velocity without any help of the resonance.[16-18] Here, we report that a single short current pulse injecting into a ferromagnetic nanodisk can determine the final direction of the core.

Figure 1 shows an atomic force microscope image of the sample and the schematic configuration used for the measurements. The samples were fabricated on sapphire ($Al_2O_3$) substrates by the lift-off method in combination with e-beam lithography. The sample consists of a Permalloy ($Fe_{19}Ni_{81}$) disk and two wide electrodes (Au(70 nm)/Cr(10 nm)), through which an electric current pulse is supplied. The thickness of the disk is 55 nm, and the diameter is 1.55 μm.

Experimental procedure is the following. First, the direction of a core magnetization is determined by magnetic force microscope (MFM) observation. Then, an electric current pulse is injected through the magnetic disk, and the core direction is again checked by MFM. Figure 2 shows the successive MFM observations with one pulse current applied between each consecutive image. The current density and the pulse duration were $1.3 \times 10^{12}$ A/m$^2$ and 2.5 ns, respectively. A dark spot at the center of the disk in Fig. 2(b) indicates that the core magnetization directs upward with respect to the paper plane. As shown in Fig. 2(c), the dark spot at the center of the disk changed into bright after the injection of the pulse current, indicating that the direction of the core magnetization has been switched. The switching occurred after every injection of the current pulse as shown in Fig. 2(b)-2(l). Thus, we can precisely control the core direction by the application of the nanosecond current pulse.

Figures 3(a)–3(e) are successive snapshots of the calculated results

for the magnetization distribution during the process of core motion and

switching induced by injection of the current pulse. The current-induced

dynamics of a vortex was calculated by micromagnetic simulations in the

framework of the Landau-Lifshitz-Gilbert (LLG) equation with a spin-

transfer term.[9, 15, 19, 20] We performed the two-dimensional calculation by

dividing the disk into rectangular prisms of $4 \times 4 \times 50$ nm$^3$; the

magnetization was assumed to be constant in each prism. The typical

material parameters for Permalloy were used: saturation magnetization $M_s$

= 1 T, exchange stiffness constant $A = 1.0 \times 10^{-11}$ J/m, and damping

constant $\alpha = 0.01$. The spin polarization of the current was taken as $P =$

0.7.[21] Noteworthy is the development of an out-of-plane magnetization

(dip) opposite to the core magnetization, which is also observed in the case of the core switching by the resonant excitation.[13, 15] The creation of a pair of the vortex and anti-vortex just before the core switching was also confirmed in the simulation. Thus, the mechanism of the switching by the pulse current can be considered to be the same as in the case of the switching by the resonant excitation.

Figure 4 shows the switching probabilities as a function of pulse duration. The switching probability is defined as the ratio of the number of the core switching events to the total trial number. The current density is $1.3 \times 10^{12}$ A/m$^2$. The high switching probabilities are obtained for the pulse duration from 2.3 to 3.0 ns, while no switching is observed for the pulse duration below 1.8 ns. Results for the pulse duration above 3 ns could not be obtained because the sample was damaged possibly due to the severe Joule heating for the longer pulse. We could not access to the higher current density above $1.3 \times 10^{12}$ A/m$^2$ due to the same reason. We confirmed that the core switching does not occur for the current density of $1.1 \times 10^{12}$ A/m$^2$ regardless of the pulse duration, indicating that the current density of $1.3 \times 10^{12}$ A/m$^2$ is just above the threshold current density.

The results of the micromagnetic simulation are superimposed on the experimental results in Fig. 4. While the simulation reproduces the existence of the threshold pulse duration below which the core is not switched, two discrepancies are found between the experimental results and the simulation; (i) The critical current density for the switching in the simulation ($2.8 \times 10^{12}$ A/m$^2$) is higher than the experimental value ($1.3 \times 10^{12}$ A/m$^2$). (ii) The critical pulse duration is longer in the experiments. Although the reason for these discrepancies is not clear at this stage, most probable are the sample heating and/or the magnetic field which are induced by the application of the current pulse.[22] The heating and the magnetic field can contribute to the reduction of the current density for the core switching in the experiments.

We have shown that a single short current pulse injecting into a ferromagnetic nanodisk can switch the direction of the core. This enables us the fast and exact control of the core direction, which is indispensable for realizing a memory cell where the bit data is stored as the direction of the nanometre-scale core magnetization. The experimentally obtained critical current density for the switching is smaller than the value in the simulation, indicating the importance of the sample heating and/or the magnetic field induced by the current pulse. Designing the cell structure to utilize the field effect in a cooperative manner to reduce the critical current is of importance from the technological point of view.

The present work was partly supported by MEXT Grants-in-Aid for Scientific Research in Priority Areas and JSPS Grants-in-Aid for Scientific Research.

**Figure captions**

Figure 1 Atomic force microscope image of the sample and schematic illustration of the experimental setup. The thickness of the disk is 55 nm, and the diameter is 1.55 $\mu$m. Two wide electrodes, through which an electric current pulse is supplied, are also seen.

Figure 2 (a) AFM image of the sample. (b) MFM image before the application of the current pulse. A dark spot at the center of the disk indicates that the core magnetization directs upward with respect to the paper plane. (c) MFM image after injecting the current pulse of $1.3 \times 10^{12}$A/m$^2$ with the pulse duration of 2.5 ns. The dark spot at the center of the disk in Fig. 2(b) changed into the bright spot, indicating the switching of the core magnetization from up to down. (d-l) Successive MFM images with a current pulse applied similarly between consecutive images. The direction of the core magnetization is switched after every injection of the current pulse.

Figure 3 Perspective view of the magnetization with a moving vortex structure. The height is proportional to the out-of-plane (z)

magnetization component. Rainbow colour indicates the in-plane component as exemplified by the white arrows in (a). (a) Initially, a vortex core magnetized upward is at rest at the disk center. (b) On application of the current pulse, the core starts to move. (c) There appears a region with downward magnetization (called 'dip' here) on the inner side of the core. (d) The dip grows as the core is accelerated. (e) After the completion of the reversal, the energy is released to a substantial amount of spin waves. Calculation with the same geometry as the experimental sample. 3D movie of micromagnetic simulation is available as "Supplementary Information".

Figure 4 Switching probabilities as a function of pulse duration. The switching probability is defined as the ratio of the number of the core switching events to the total trial number, which ranges between 20 and 80 times for each duration. The big symbols represent the experimental data, where the blue and red ones correspond to the case for the current densities of $1.1 \times 10^{12}$ A/m$^2$ and $1.3 \times 10^{12}$ A/m$^2$,

respectively. The green dotted line is the simulation results with the current density of $2.8 \times 10^{12} A/m^2$.

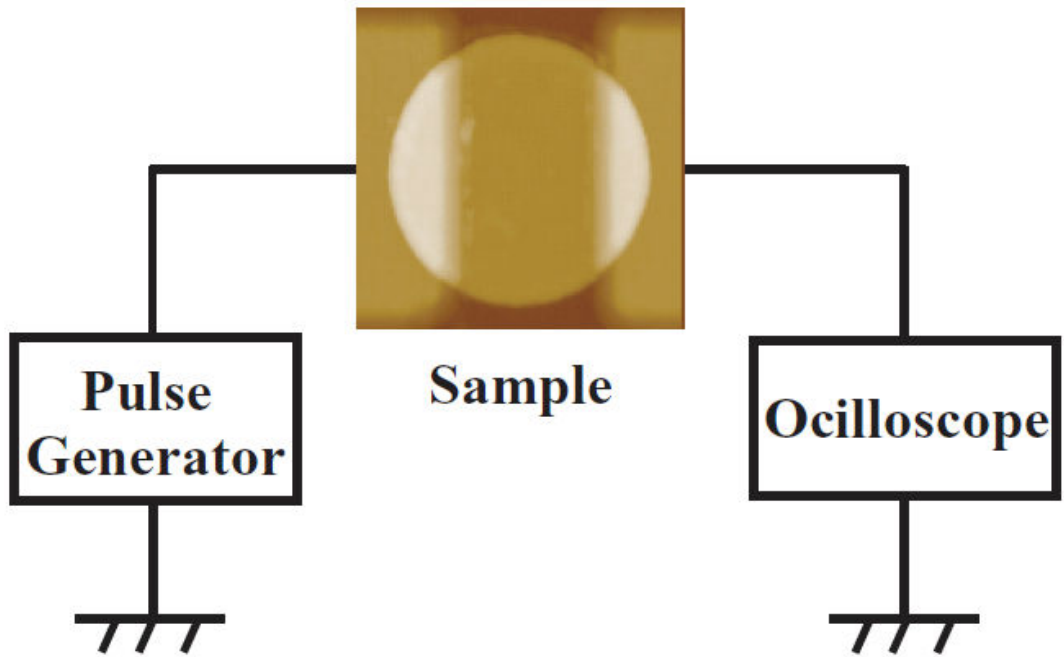

Fig. 1

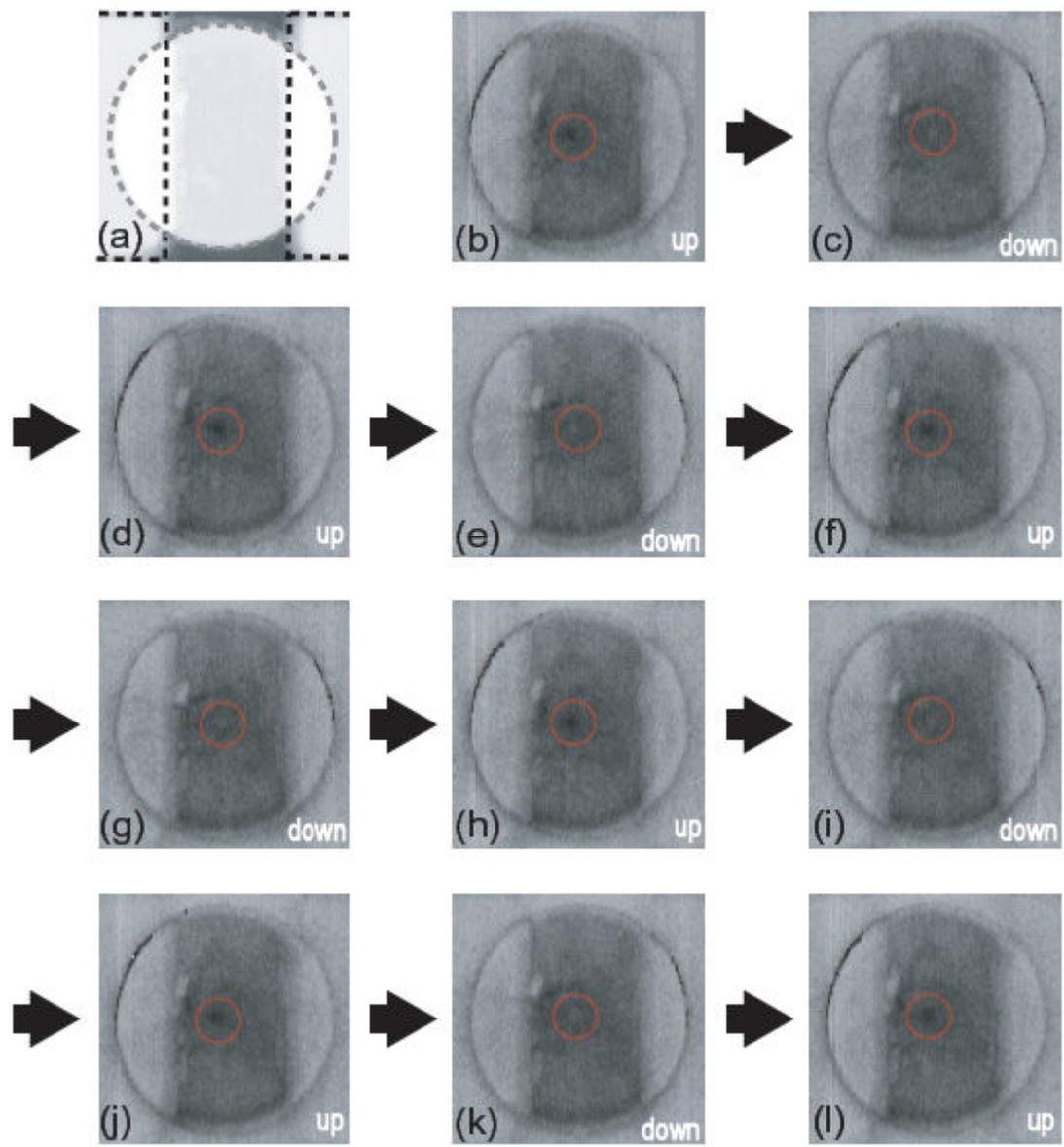

Fig. 2

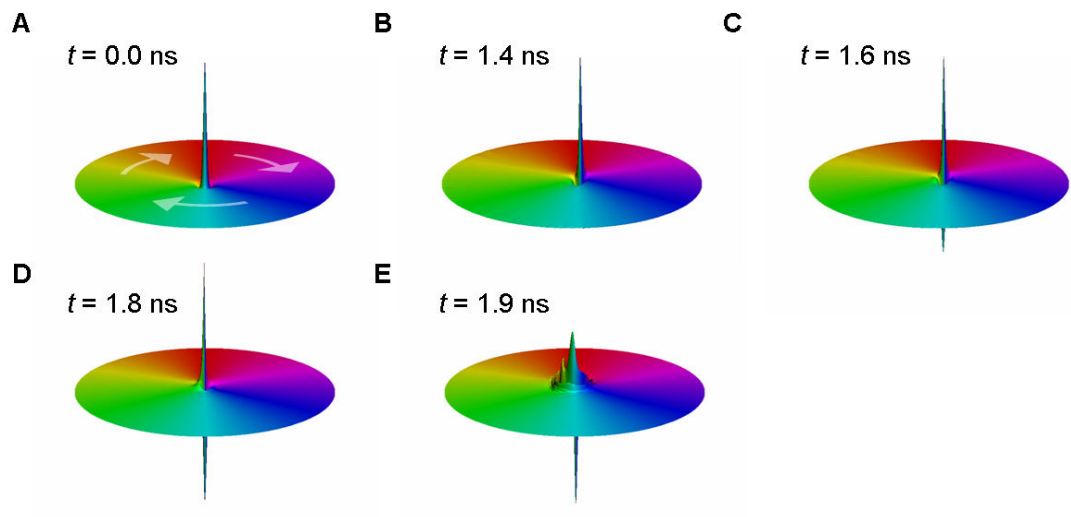

Fig. 3

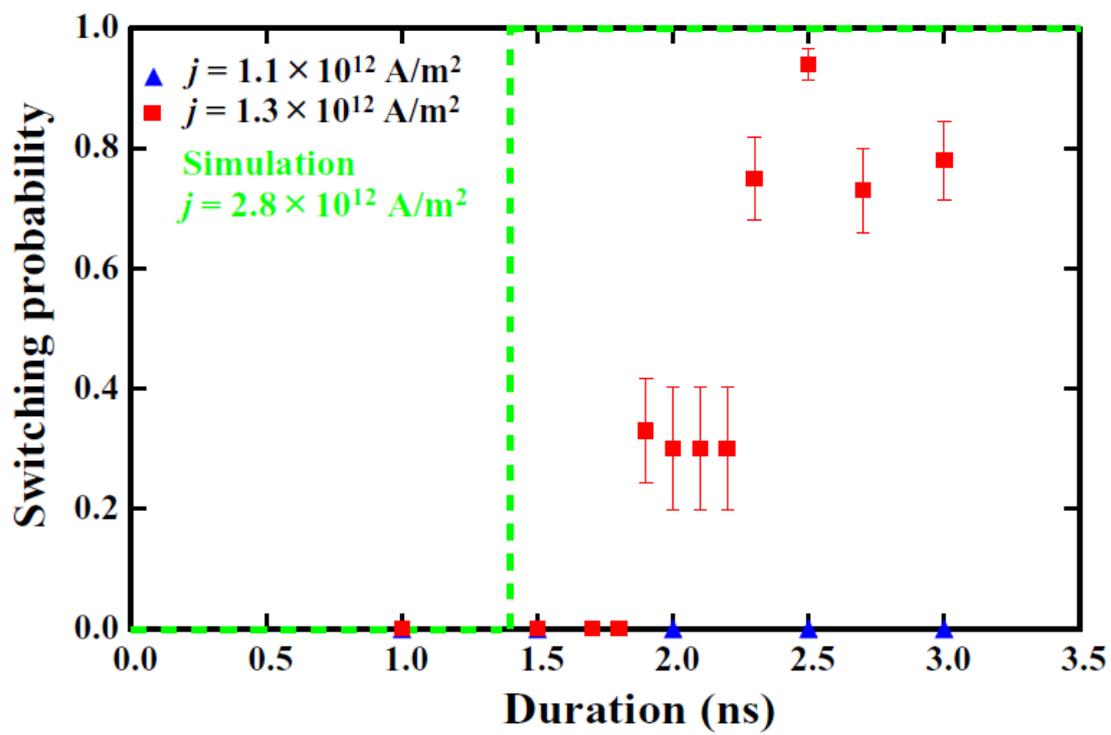

Fig. 4